\documentclass[aps,pra,preprint,groupedaddress,showpacs]{revtex4-1}
\usepackage{amssymb}
\usepackage{amsmath}
\usepackage{dcolumn}
\usepackage{bm}
\usepackage{graphicx}
\usepackage{mathrsfs}

\begin{document}

\title{Hanbury-Brown and Twiss effect without quantum interference in photon counting regime}
\author{Bin Bai $^{1}$}
\author{Yu Zhou $^{1,2*}$}
\author{Hui Chen $^{1}$}
\author{Huai bin Zheng $^{1}$}
\author{Jian bin Liu $^{1}$}
\author{Rui feng Liu $^{2}$}
\author{Yun long Wang $^{2}$}
\author{Zhuo Xu $^{1}$}
\author{Fuli Li $^{2}$}

\address{$^1$Electronic Materials Research Laboratory, Key Laboratory of the Ministry of Education \& International Center for Dielectric Research, Xi'an Jiaotong University, Xi'an 710049, China}
\address{$^2$MOE Key Laboratory for Nonequilibrium Synthesis and Modulation of Condensed Matter, and Department of Applied Physics, Xi'an Jiaotong University, Xi'an 710049, China}

\address{$^*$Corresponding author: zhou1@mail.xjtu.edu.cn}

\begin{abstract}
Usually HBT effect can be interpreted by classical (intensity fluctuation correlation) and quantum (interference of two-photon probability amplitudes) theories properly at the same time. In this manuscript, we report a deliberately designed experiment in which two chaotic light beams has the same intensity fluctuation but mutual-orthogonal polarizations to each other so there will be no interference of two-photon probability amplitudes. Classical and quantum theory give different predictions on if there should be HBT (photon bunching) effect or not in the experiment. The experiment results are used to test the two different predictions. At the end, both the temporal and spatial HBT effects are observed.
\end{abstract}

\maketitle

\section{introduction}
In 1956, Hanbury Brown and Twiss (HBT) demonstrated a surprised phenomenon of the light \cite{hbt,hbt2}. In their historic experiment two photo-detectors are placed in the far field zone of a chaotic radiation source. The correlation between the signals from the two detectors is measured. The intensities from each detector are constant when the detector is scanned along the transverse direction. However, if one detector is fixed and the other is scanned, the correlation between them is not a constant. A peak would be observed in a certain position. This peak is approximately twice bigger than the product of the single intensities from two detectors. This experiment typifies all subsequent measurements of the second-order correlation. This effect is called HBT effect ever since.

For HBT effect, there are two different interpretations---one from classical and one from quantum theory.
In classical theory, HBT is interpreted as the result of the correlation of intensity fluctuations \cite{hbt3}. In the HBT experiment a beam splitter splits the light from a chaotic source and makes two identical copies of light field in different places. When two detectors are placed in the same coherence volumes of the light fields--- even the two coherence volumes are spatially separated due to the beam splitter, a correlation between the two detection results can be found \cite{mandel}. This correlated detection is ascribed to the correlation of  intensity fluctuations in the two identical coherence volumes of chaotic light field. When two detectors are placed in different coherence volumes, there are no correlated intensity fluctuations for each detectors and no correlation is detected.

%A peak is observed when radiation of the same mode reaches the two detectors. When light from different modes reach those two detectors, the result of correlation equals to the product of the individual intensities from each detector. The radiation field from a chaotic source is considered as a random process and every mode is independent. If the same radiation mode hits two detectors, the fluctuations of signals from the two detectors would be correlated.

%When detectors record different modes from radiation, there is no peak. And then Glauber described the process of photodetection by quantum mechanical means and thought that the statistical correlation is between the photon number fluctuations \cite{glauber}.

In quantum theory, a full quantum mechanical interpretation of HBT experiment applies the ``law of combing amplitudes"---the coherent superposition of different but indistinguishable two-photon probability amplitudes \cite{feyman,fano,wu,shih,liu}. There are two photons(i and j)from the chaotic source and two detectors(A and B) for their detections. There are two different ways to trigger a joint detection event---both detectors are triggered by photons. One way is that photon i goes to detector A and photon j goes to detector B, in the language of quantum mechanics this is the two-photon probability amplitudes A$_I$. Another way is that photon j goes to detector A and photon i goes to detector B, we call it probability amplitude A$_{II}$. When the two different probability amplitudes are indistinguishable, the probability of the joint photons detection is $P_{cc}=\mid A_I+A_{II}\mid^2$. When the light is chaotic the interference term gives the extra probability of detecting joint events than that if the light source is a coherent source say, a laser. So in quantum point of view, the peak observed in HBT effect comes from the interference between different but indistinguishable two-photon probability amplitudes \cite{twophoton1,twophoton2,twophoton3}. The prerequisite of the happening of quantum interference is that the indistinguishable of probability amplitudes. If there is a method to distinguish two probability amplitudes, even in principle, there will be no quantum interference and no HBT effect will be observed \cite{mandel}.

In most HBT type experiments, both classical and quantum theory give the same predictions and explain experimental results properly. In this manuscript, we report a specially designed experiment in which two theories give different predictions and test their predictions with experimental results. In our experiment, two pseudo-thermal light beams from different sources have the same intensity fluctuations and orthogonal polarizations to each other. According to classical theory, because the two beams have the same intensity fluctuation we should observe HBT effect. One the other hand, because photons from two light beams are distinguishable (their polarizations are orthogonal to each other) there should be no interference between two-photon probability amplitudes and we should not observe HBT effect---the bunching of photons. In the end of our experiments, both spatial and temporal HBT effect are observed.

This paper is organized as follows: we will first briefly present two different theories about the correlation of intensity fluctuations and two-photon interference in the Sec. $\mathrm{\uppercase\expandafter{\romannumeral2}}$. Experiments and discussion are in Sec. $\mathrm{\uppercase\expandafter{\romannumeral3}}$. Section $\mathrm{\uppercase\expandafter{\romannumeral4}}$ summaries the
conclusions.
\section{Theory}

In classical theory, HBT is interpreted as the result of the correlation of intensity fluctuations. HBT experiment measures the correlation between the output of two photo-detectors located in two different space-time points $(t_{A},\overrightarrow{r}_{A})$ and $(t_{B},\overrightarrow{r}_{B})$. The quantity is the second-order coherence function:

 \begin{eqnarray}
G^{(2)}(t_{A},\overrightarrow{r}_{A};t_{B},\overrightarrow{r}_{B})&=&{\left \langle I_{A}(t_{A},\overrightarrow{r}_{A})I_{B}(t_{B},\overrightarrow{r}_{B})\right \rangle}\nonumber\\
&=&{\left \langle E^{*}_{A}(t_{A},\overrightarrow{r_{A}})E_{A}(t_{A},\overrightarrow{r_{A}})E^{*}_{B}(t_{B},\overrightarrow{r_{B}})E_{B}(t_{B},\overrightarrow{r_{B}})\right \rangle},
\end{eqnarray}
where $I_{\emph{i}}(t_{\emph{i}},\overrightarrow{r_{\emph{i}}})$ and $E_{\emph{i}}(t_{\emph{i}},\overrightarrow{r}_{\emph{i}})$, $\emph{i}=(A,B)$, are intensities and electric fields at each detector, respectively. For a chaotic radiation, we can realize that the radiation is the sum of the contribution of many microscopic sources. We can write $E_{A}(t_{A},\overrightarrow{r}_{A})$ and $E_{B}(t_{B},\overrightarrow{r}_{B})$ as a discrete sum of N components:

\begin{eqnarray}
  E_{A}(t_{A},\overrightarrow{r}_{A}) &=& \sum_{\emph{j}}^{N}E_{A\emph{j}}(t_{A\emph{j}},\overrightarrow{r}_{A\emph{j}}), \nonumber\\
  E_{B}(t_{B},\overrightarrow{r}_{B}) &=& \sum_{\emph{j}}^{N}E_{B\emph{j}}(t_{B\emph{j}},\overrightarrow{r}_{B\emph{j}}),
\end{eqnarray}
where $A\emph{j}$ indicates that the radiation arrives at the detector A from the $\emph{j}$ -th element of the source. The phases of the electric field from each microscopic source is independent and random. As a result some terms will vanish when the ensemble average is calculated. The function can be simplified as follows:

\begin{eqnarray}
G^{(2)}(t_{A},\overrightarrow{r}_{A};t_{B},\overrightarrow{r}_{B})&=&{\left \langle I_{A}(t_{A},\overrightarrow{r}_{A})\right \rangle}{\left \langle I_{B}(t_{B},\overrightarrow{r}_{B})\right \rangle}+ \left |  \Gamma_{12}^{(1)}(t_{A},\overrightarrow{r}_{A};t_{B},\overrightarrow{r}_{B})\right |^{2}\nonumber\\
&=&{\left \langle I_{A}\right \rangle}{\left \langle I_{B}\right \rangle}\left [ 1+  \left |  \gamma(t_{A},\overrightarrow{r}_{A};t_{B},\overrightarrow{r}_{B})\right |^{2}\right ],
\end{eqnarray}
where $\left \langle I_{A}\right \rangle$ and $\left \langle I_{B}\right \rangle$ are the average intensities recorded by A detector and B detector; $\Gamma_{12}^{(1)}$ is the mutual coherence function and $\gamma$ is the first degree of coherence.

The concept of intensity fluctuations is defined as:
\begin{eqnarray}
\Delta I_{\emph{i}}= I_{\emph{i}}-\left \langle I_{\emph{i}}\right \rangle,
\end{eqnarray}

The correlation between intensity fluctuations is mathematically expressed :

\begin{eqnarray}
\left \langle \Delta I_{A}\Delta I_{B} \right \rangle &=&\left \langle  (I_{A}-\left \langle I_{A}\right \rangle)(I_{B}-\left \langle I_{B}\right \rangle) \right \rangle\nonumber\\
&=&{\left \langle I_{A}I_{B}\right \rangle}-{\left \langle I_{A}\right \rangle}{\left \langle I_{B}\right \rangle},
\end{eqnarray}

Comparing this expression to the above equation, it is realized that HBT effect is due to the correlation of the intensity fluctuations of the radiation at two detectors. When the intensity fluctuations recorded at two photo-detectors are same,  the peak appears and the HBT effect can be observed in the classical interpretation.

From another point of view, let us see the interpretation of two-photon interference. This interpretation is mainly the interference of indistinguishable two-photon probability amplitudes. For the second-order phenomena, the quantity which is measured is the probability of jointly producing two photo-electron events at space time points $(t_{A},\overrightarrow{r}_{A})$ and $(t_{B},\overrightarrow{r}_{B})$. From the second-order Glauber correlation function \cite{glauber,glauber2,glauber3}, $G^{(2)}$ is as follow:

\begin{eqnarray}
G^{(2)}(t_{A},\overrightarrow{r}_{A};t_{B},\overrightarrow{r}_{B})={\left \langle E_{A}^{(-)}(t_{A},\overrightarrow{r}_{A})E_{B}^{(-)}(t_{B},\overrightarrow{r}_{A})E_{B}^{(+)}(t_{B},\overrightarrow{r}_{B})E_{A}^{(+)}(t_{A},\overrightarrow{r}_{A})\right \rangle}
\end{eqnarray}

The photons come from the chaotic source, so a pure state composed by two independent photons is described as \cite{qo}:

 \begin{eqnarray}
\left|\Psi_{\emph{i,j}}\right\rangle&=&\left|\Psi_{\emph{i}}\right\rangle\left|\Psi_{\emph{j}}\right\rangle\nonumber\\&=& A\int \emph{d}\omega \emph{f}(\omega)\emph{e}^{-\emph{i}\omega t_{0i}}a^{\dag}(\omega)\left|0\right\rangle\int \emph{d}\omega' \emph{f}(\omega')\emph{e}^{-\emph{i}\omega' t_{0j}}a^{\dag}(\omega')\left|0\right\rangle
\end{eqnarray}

They are composed by two independent single photon wavepackets. The time of creation of two photons is distinguishable. The second-order correlation function is as following:

 \begin{eqnarray}
G^{(2)}(t_{A},\overrightarrow{r}_{A};t_{B},\overrightarrow{r}_{B})&=&{\left \langle \Psi_{\emph{i,j}}\right | E_{A}^{(-)}(t_{A},\overrightarrow{r}_{A})E_{B}^{(-)}(t_{B},\overrightarrow{r}_{A})E_{B}^{(+)}(t_{B},\overrightarrow{r}_{B})E_{A}^{(+)}(t_{A},\overrightarrow{r}_{A})\left|\Psi_{\emph{i,j}}\right\rangle}\nonumber\\
&=& \left | {\left \langle 0 \right | E_{B}^{(+)}(t_{B},\overrightarrow{r}_{B})E_{A}^{(+)}(t_{A},\overrightarrow{r}_{A})\left|\Psi_{\emph{i,j}}\right\rangle} \right |^{2}\nonumber\\&\equiv& \left | \Psi_{i,j}(t_{A},\overrightarrow{r}_{A};t_{B},\overrightarrow{r}_{B}) \right |^{2}
\end{eqnarray}

In the HBT experiment, photons trigger two detectors ($D_{A}$ and $D_{B}$). The function is calculated to be:

\begin{eqnarray}
G^{(2)}(t_{A},\overrightarrow{r}_{A};t_{B},\overrightarrow{r}_{B})&=&\left | {\left \langle 0 \right | E_{B}^{(+)} \left |\Psi_{\emph{i}} \right \rangle}{\left \langle 0 \right | E_{A}^{(+)}\left |\Psi_{\emph{j}}\right \rangle}+{\left \langle 0 \right | E_{B}^{(+)}\left |\Psi_{\emph{j}}\right \rangle}{\left \langle 0 \right | E_{A}^{(+)}\left |\Psi_{\emph{i}}\right \rangle} \right |^{2}\nonumber\\&=&\left | \mathbf{A}_{i\rightarrow A;j\rightarrow B}(t_{A},\overrightarrow{r}_{A};t_{B},\overrightarrow{r}_{B})+\mathbf{A}_{i\rightarrow B;j\rightarrow A}(t_{A},\overrightarrow{r}_{A};t_{B},\overrightarrow{r}_{B}) \right |^{2}
\end{eqnarray}

The physics of two-photon interference phenomenon is shown in this equation. Two quantities:

\begin{eqnarray}
\textbf{A}_{1}(t_{A},\overrightarrow{r}_{A};t_{B},\overrightarrow{r}_{B})&=&\textbf{A}_{i\rightarrow A;j\rightarrow B}\nonumber\\ &=& {\left \langle 0 \right | E_{A}^{(+)} \left |\Psi_{\emph{i}} \right \rangle}{\left \langle 0 \right | E_{B}^{(+)} \left |\Psi_{\emph{j}} \right \rangle}\nonumber\\ &=&\Psi_{iA}(t_{A},\overrightarrow{r}_{A})\Psi_{jB}(t_{B},\overrightarrow{r}_{B})
\end{eqnarray}

\begin{eqnarray}
\textbf{A}_{2}(t_{A},\overrightarrow{r}_{A};t_{B},\overrightarrow{r}_{B})&=&\textbf{A}_{i\rightarrow B;j\rightarrow A}\nonumber\\ &=& {\left \langle 0 \right | E_{B}^{(+)} \left |\Psi_{\emph{i}} \right \rangle}{\left \langle 0 \left | E_{A}^{(+)}\right |\Psi_{\emph{j}}\right \rangle}\nonumber\\ &=&\Psi_{iB}(t_{A},\overrightarrow{r}_{A})\Psi_{jA}(t_{B},\overrightarrow{r}_{B})
\end{eqnarray}
are two-photon amplitudes. They are indistinguishable. $\mathbf{A}_{1}$ expresses that photon $i$ is recorded by detector $D_{A}$ and photon $j$ is recorded by detector $D_{B}$. At the same time, $\mathbf{A}_{2}$ expresses that photon $i$ is recorded by detector $D_{B}$ and photon $j$ is recorded by detector $D_{A}$.

The calculation of the second-order correlation function is written as:

\begin{eqnarray}
G^{(2)}(t_{A},\overrightarrow{r}_{A};t_{B},\overrightarrow{r}_{B})&\propto&\left | \textbf{A}_{1}(t_{A},\overrightarrow{r}_{A};t_{B},\overrightarrow{r}_{B}) \right |^{2}+\left | \textbf{A}_{2}(t_{A},\overrightarrow{r}_{A};t_{B},\overrightarrow{r}_{B}) \right |^{2}\nonumber\\&+&2\emph{Re}\textbf{A}_{2}^{*}(t_{A},\overrightarrow{r}_{A};t_{B},\overrightarrow{r}_{B})\textbf{A}_{1}(t_{A},\overrightarrow{r}_{A};t_{B},\overrightarrow{r}_{B})
\end{eqnarray}

It shows that the HBT effect is observed because of the superposition of amplitudes. The law of combing amplitudes is used in the joint photo-detection event. The superposition takes place between two alternative, different and indistinguishable amplitudes.

\section{Experiment and Discussion}

\begin{figure}[htb]
    \centering
    \includegraphics[width=160mm]{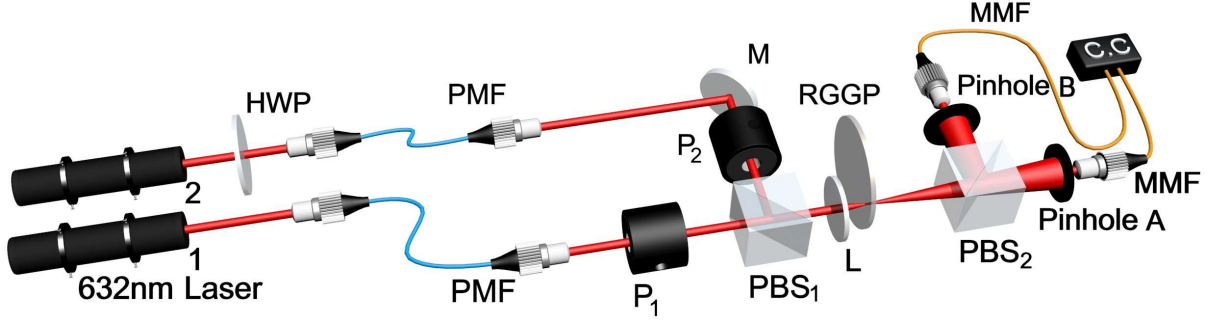}
    \caption{The scheme of experiment. The sources are both He-Ne lasers with the light wavelength of $632.8 nm$. HWP is a half wave plate which changes the polarization of light from laser 2. Two PWF are two single mode polarization-maintaining fibers. $P_1$ and $P_2$ are two polarizers which are used to keep the polarizations of two beams unchanged. $PBS_1$ and $PBS_2$ are two polarization beam splitters. Two beams are combined into one beam at $PBS_1$ and the combined beam is focus by the lens (L) on the rotating ground glass plate (RGGP). $PBS_2$ splits the combined beam into two beams of light. Two MMF are multi-mode fibers. They transfer light to two single-photon detectors. It is measure by a standard HBT intensity interferometer.}
    \label{Exp}
\end{figure}

The experimental setup is shown in Fig. \ref{Exp}. In the experiment we use two He-Ne lasers with wavelength at $632.8$ nm. The polarizations of both lasers are horizontal initially. There is a half wave plate behind laser $2$ which turns the polarization of the light beam from laser $2$ into vertical, so the polarizations of two laser light beams behind the half wave plate become orthogonal to each other. The two beams pass through two single mode polarization-maintaining fibers and two polarizers $P_1$ and $P_2$. $P_1$ is set to horizontal polarization and $P_2$ is set to vertical polarization respectively to keep the polarizations of two beams unchanged.

\begin{figure}
     \centering
    \includegraphics[width=80mm]{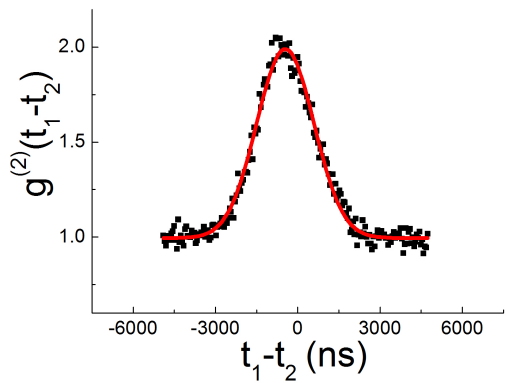}
   \caption{The result is shown when only one laser works. Transmission path for the first polarization beam splitter exists. The other beam of light doesn't exist because of one laser doesn't work. The constant of the peak is about 33\%. }
   \label{Ti}
\end{figure}

In the first step of the experiment, our HBT intensity interferometer is tested. In this step, we only turn laser $1$ on and keep laser $2$ off. The beam is focused by a lens (L) on a rotating round ground glass plate (RGGP) to generate pseudo-thermal light. Because the generated pseudo-thermal light is also horizontal polarized, the light can only reach detector B and there is no coincidence counts found in our measurement. Then an additional half wave plate is placed between the first polarized beam splitter ($PBS_1$) and the rotating ground glass (not shown in Fig. \ref{Exp}). It turns the horizontal polarized laser beam from laser $1$ into $45^{\circ}$ with respect to horizontal polarization. Because the polarization of the light beam is not completely vertical or horizontal, it is split into two beams of light after it passes through the second polarization beam splitter ($PBS_2$). Two detectors would both be triggered by photons and it is measured by a standard HBT intensity interferometer. The result is shown in the Fig. \ref{Ti}. The FWHM of the peak which is about 2.59 $\mu m$ and determined by the rotating speed of the ground glass. It shows that the HBT effect exists when $45^{\circ}$ polarized light beam through rotating ground glass and the polarization beam splitter. From this step we can see that our HBT intensity interferometer works properly. The same test is repeated by turning laser $2$ on and keeping laser $1$ off.

In above test, when one light beam (from laser $1$ or $2$) with polarization set to $45^{\circ}$ passes through the rotating ground glass and the polarization beam splitter, there are two sets of identical speckle patterns. The detectors would record the same intensity fluctuations or the photon number fluctuations as long as they are in symmetric positions. Therefore, the observed HBT effect can be interpreted as the correlation of the intensity fluctuations of the light field at the positions of two detectors (classical interpretation). On the other hand, one beam of light is split into two by $PBS_2$ because the polarization of light is not completely vertical or horizontal. There are two different but indistinguishable probability amplitudes for the joint photo-detection event. The photon bunching effect can be interpreted as the result of the interference between two probability amplitudes.
Quantum theory can also explain the experimental result.

In the second step of experiment, laser 1 and laser 2 are both turned on at the same time. The polarization of the light beam from laser 1 is horizontal and the polarization from laser 2 is vertical. The two beams are combined into one beam at $PBS_1$. The combined beam is focus by the lens (L) on the rotating ground glass plate (RGGP). However, the areas where two beams of light are focused on the ground glass are slightly different with each other, which is easy to make. Now we have two sets of different speckles: the first set of speckles is from laser 1 with horizontal polarization and can only reach detector A; the second set of speckles is from laser 2 with vertical polarization and can only reach detector B. Two sets of speckles are different in the distribution of intensity fluctuation because they are generated from different areas on the RGGP. The combined beam passes through the second polarized beam splitter $PBS_2$ and is split into two beams of light because of their mutual-orthogonal polarization. Detector A records photons from laser 1 with horizontal polarization and detector B records photons from laser 2 with vertical polarization, respectively.

\begin{figure}
     \centering
    \includegraphics[width=80mm]{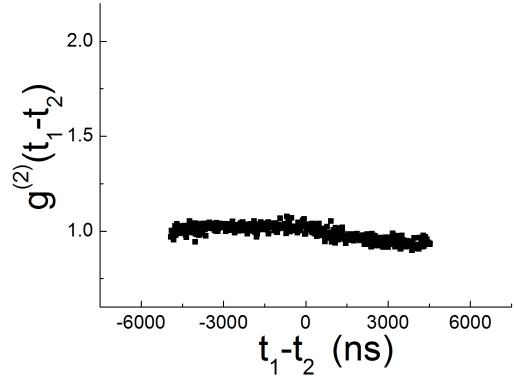}
    \caption{The result is shown when two laser both open and speckle is different. The peak of correlation does not appear. The result is fiat.}
    \label{Di}
\end{figure}

The result is shown in Fig. \ref{Di}. The $G^{(2)}$ function is flat and no bunching effect is observed. Both classical and quantum theories can explain the result properly. For classical theory, the speckles of two beams of light are different so their intensity fluctuations are different. When it is measured by the standard HBT intensity interferometer, the peak can not appear because intensity fluctuations recorded at detector A and B are different. At the same time, from quantum point of view the photons recorded by two detectors are distinguishable because of their mutual-orthogonal polarizations. The photon from laser $1$ with horizontal polarization can only trigger detector A and The photon from laser $2$ with vertical polarization can only trigger detector B. According to the theory of two-photon interference, the photon bunching effect also can not appear because there is only one two-photon probability amplitude left in the experiment--there is no chance for quantum interference with only one probability amplitude left. As before, both theories can explain the result properly.

In the third step of our experiment, the alignment is the same as the second step except that we carefully align the two beams to be focused (by lens (L)) \emph{on the same area} of the rotating ground glass plate (RGGP) to generate two sets of pseudo-thermal light with identical spatial and temporal intensity fluctuation distribution but mutual-orthogonal polarizations. The pseudo-thermal light passes through $PBS_2$ and is measured by the standard HBT intensity interferometer. Light from laser $1$ passes through two beam splitters and only triggers detector A due to its horizontal polarization. For the same reason, light from laser $2$ is reflected by $PBS_2$ and only triggers detector B.

\begin{figure}
    \centering
    \includegraphics[width=80mm]{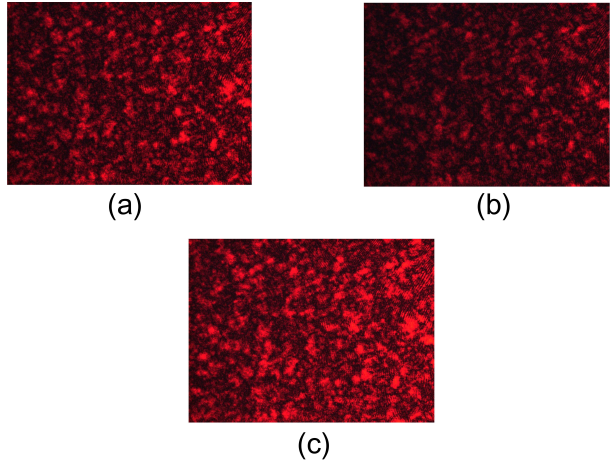}
    \caption{ The speckle patterns are shown in three pictures. Picture (a) is the speckle pattern of the reflected light with horizontal polarization through the first polarization beam splitter. At the same time Picture (b) shows the pattern of the transmitted light with vertical polarization. Picture (c) is the image of the speckle pattern when both of two beams of light pass the first polarization beam splitter at the same time. Though the intensity of light in (a) and (b) is not equal, they  have almost the same speckle pattern. }
   \label{sanban}
\end{figure}

In this setup, classical and quantum theory will give different predictions on if HBT effect (bunching effect) will be observed. From classical point of view, HBT effect will be observed: we carefully make the focused horizontal-polarized (from laser 1 only) and vertical-polarized (from laser 2 only) laser beam on the same area of the ground glass plate. This alignment makes sure the generation of two sets of identical speckle patterns while their polarizations are orthogonal to each other. This is verified by placing a CCD after the $RGGP$. We check speckle patterns generated by blocking each laser beams in turn and find them almost identical, as shown in Fig. \ref{sanban}. We notice that after $PBS_{2}$ the vertical-polarized light are reflected and goes to detector B and the horizontal-polarized light are transmitted and goes to detector A only. Since the two sets of pseudo-thermal light have identical intensity fluctuations (when the ground glass is rotating), HBT effect is expected to be observed according to classical theory (intensity fluctuation correlation). On the other hand, quantum theory (two-photon probability amplitudes interference)  predicts there will be no HBT effect observed: the photons from laser $1$ only go to detector A and photons from laser $2$ only go to detector B. There is no chance that photons from laser $1$ go to detector B or photons from laser $2$ go to detector A because their polarizations and the $PBS_{2}$ does not allow that happens. We check this by blocking each beams in front of $PBS_{1}$ in turn and the coincidence counts drop to zero instantly. So all coincidence events must come from one photon (with horizontal polarization) from laser 1 to trigger detector A and one photon (with vertical polarization) from laser 2 to trigger detector B. Under this circumstance, two-photon probability amplitudes interference is impossible because there is only one probability amplitude left.

\begin{figure}
  \centering
  \includegraphics[width=80mm]{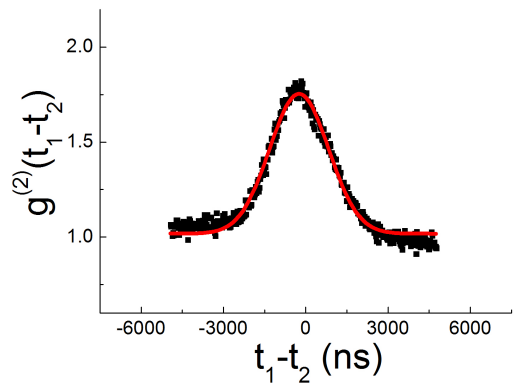}
  \caption{The result is shown when two laser both open and the speckle patterns of two beams of light are same. The peak appears. The peak width at half-height is 2800ns. The constant of the peak is 27\%.}
  \label{Sa}
\end{figure}

\begin{figure}
   \centering
  \includegraphics[width=100mm]{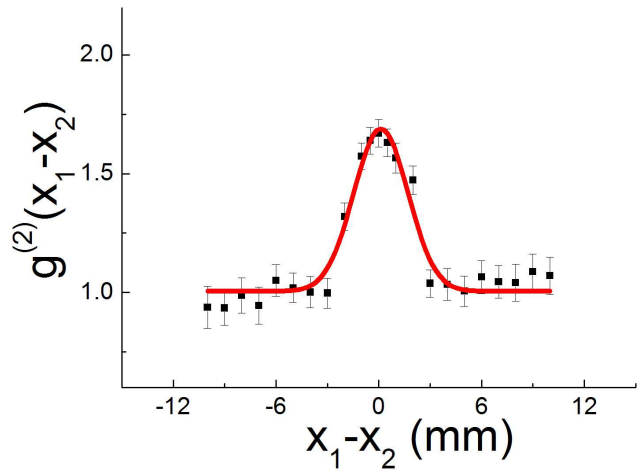}
  \caption{The result is shown when a detector is fixed and the other detector moves horizontally. The peak width at half-height is 3.6 mm. The visibility of the peak is about 25\%.}
  \label{Lo}
\end{figure}

The result of this experiment is shown in Fig. \ref{Sa}. Firstly, we measure the $G^{(2)}$ function in temporal domain. It shows that a peak appears which means that there is a temporal correlation between the light field at points of detectors A and B. The temporal HBT effect is be observed. Then we measure the $G^{(2)}$ function in spatial domain.  We measure the coincidence counting rate when one of the detectors moves horizontally and the other detector is fixed, the result is in Fig. \ref{Lo}. The spatial HBT effect is also observed.

\section{Conclusion}
With classical (intensity fluctuation correlation) and quantum (interference of two-photon probability amplitudes) theories the HBT experiment usually can be explained properly at the same time. The classical theory emphasizes the correlation between the same intensity fluctuations. The quantum theory of interference of two-photon probability amplitude emphasizes that the probability amplitudes are indistinguishable. In this manuscript, an experiment in which two chaotic light beams have the same intensity fluctuation distribution but mutual-orthogonal polarizations to each other is designed. In this setup, two theories gives two different predictions on if HBT effect will be observed or not. The experimental results show that both the temporal and spatial HBT effects are observed. This result can be explained by the theory of the intensity fluctuation but not by the theory of interference of two-photon probability amplitudes.

 Two sets of pseudo-thermal light have identical intensity fluctuation distributions, so HBT effect should be observed according to classical theory. However, since their polarizations are mutual-orthogonal, the photons from laser 1 only go to detector A and photons from laser 2 only go to detector B. So there is only one probability amplitude left and interference between two-photon probability amplitudes can not happen. The HBT effect observed in the third step of our experiment described in this manusript can not be explained by the quantum (interference of two-photon probability amplitudes) theory.

We noticed that in recent research the correlation of HBT effect in ghost imaging is separated into two parts--the classical part and the quantum part on the criteria of quantum discord \cite{SR}. According to the  paper mentioned, the quantum correlation and classical correlation does exist in any intensity of light.  When the light is very weak, the quantum part is bigger than classical part. With the increase of light intensity, the quantity of classical correlation will exceed that of quantum correlation. In our experiment, the correlation of HBT is recognized as classical one because the quantum interpretation (two-photon interference interpretation)is ruled out by our setup. This does not mean there is no quantum correlation in the observed HBT effect on the criteria of quantum discord.

\section*{Funding Information}
This work is supported by the National Basic Research Program of China (973 Program) under Grant No.2015CB654602 and the 111 Project of China under Grant No. B14040.

\end{document}